\documentclass{aa}
\usepackage{psfig} 

\def\ros{{\sl ROSAT }}
\def\asca{{\sl ASCA }}

\def\ein{{\sl Einstein }}

\def\G{$\Gamma_{\rm x}$ }
\def\NH{$N_{\rm H}$ }

\def\0134{RX\,J0134-4258}     
\def\7{QSO\,0117-2837}     
\def\4c{4C\,+74.26} 
 
\def\approxlt{\mathrel{\hbox{\rlap{\lower.55ex \hbox {$\sim$}}
        \kern-.3em \raise.4ex \hbox{$<$}}}}
\def\approxgt{\mathrel{\hbox{\rlap{\lower.55ex \hbox {$\sim$}}
        \kern-.3em \raise.4ex \hbox{$>$}}}}

\begin{document}

  \thesaurus{03         
             (11.01.2;  
               11.14.1;  
               11.17.3;  
               11.19.1;  
               13.25.2)  
}

 \title{X-ray emission/absorption mechanisms of 4 NLSy-1-like AGN and a radio quasar } 
\subtitle {{ QSO 0117-2837, RX\,J0134.3-4258, NGC\,4051, Mrk\,1298, 4C\,+74.26}}  
 \author{Stefanie Komossa\inst{1}, Janek Meerschweinchen\inst{2}}     
\offprints{St. Komossa,
 skomossa@xray.mpe.mpg.de}
\institute{
Max-Planck-Institut f\"ur extraterrestrische Physik,
         Postfach 1603, D-85740 Garching, Germany 
 \and 
 Weststrasse 19, D-3063 Obernkirchen 2, Germany }
\date{Received: April 1999; accepted: 5 November 1999}
   \maketitle
\markboth{St. Komossa \& J. Meerschweinchen, X-ray properties of four NLSy1s and a radio quasar} 
{St. Komossa \& J. Meerschweinchen, X-ray properties of four NLSy1s and a radio quasar}

   \begin{abstract}
We present a study of the X-ray variability properties 
and spectral shapes of five active galaxies
all of which show extreme or enigmatic X-ray properties. 
We focus on \7, \0134, and NGC\,4051, and 
briefly comment on Mrk\,1298 and \4c.
The individual objects were originally partly selected as candidates to host 
warm absorbers on the basis of (i) characteristic X-ray absorption features
(NGC\,4051, Mrk\,1298, \4c), 
(ii) extreme X-ray spectral steepness (\7; this object is found
to be located in the `zone of avoidance' when plotted in
the $\Gamma_{\rm x}$--FWHM$_{\rm H\beta}$ diagram), and (iii) drastic spectral
variability (\0134).
The temporal analysis reveals 
large-amplitude variability by a factor $\sim$30 in the 
long-term X-ray lightcurve of NGC\,4051, 
very rapid variability of Mrk\,1298, constant X-ray flux 
of the NLSy1 galaxy \7, and constant mean countrate of 
\0134 despite huge spectral changes.  
Besides the warm absorber, 
several further mechanisms     
and their merits/shortcomings
are investigated to explain the spectral 
characteristics of the individual objects. Different models
are favored for different sources.
Consequences for Narrow-line Seyfert 1s in general are discussed
and we present results from photoionization 
models to distinguish between different suggested NLSy1 scenarios. 
 
   \end{abstract}
%
\section{Introduction}

\subsection {Narrow-line Seyfert\,1 galaxies}

X-ray and optical observations of the last decade revealed 
a new sub-class of active galaxies that shows a number
of unusual properties which are still not well understood. 
The subgroup of Narrow-line Seyfert\,1 galaxies (NLSy1s hereafter)
was recognized by Osterbrock \& Pogge (1985) based on optical
properties, namely, the small widths of the lines emitted from
the broad line region (BLR). 
Puchnarewicz et al. (1992) made the interesting observation 
that many optical spectra of a sample of ultrasoft X-ray AGN 
discovered during the \ein survey turned out to be NLSy1s, confirming
the suggestion of Stephens (1989) that `X-ray selection may be an
efficient way to find NLSy1 galaxies'.  
Many more galaxies of this type were identified in the course
of optical follow-up observations of \ros X-ray sources 
(e.g., Bade et al. 1995, Greiner et al. 1996, Becker et al. 1996,
Moran et al. 1996, Wisotzki \& Bade 1997, Grupe et al. 1998, 
Xu et al. 1999).

Correlation analyses performed in the last few years confirmed and 
quantified the trend that was already present in the 
study of Puchnarewicz et al. (1992): the correlation of 
{\em steep} X-ray spectra (measured at soft X-ray energies) 
with {\em small} widths of the BLR Balmer lines (e.g.,  
Laor et al. 1994, 1997,
Boller et al. 1996 (BBF96 hereafter), Brandt et al. 1997,
Grupe et al. 1999a).{\footnote {
More precisely, the clearest trend is that {\em large}-FWHM BLR lines
always appear in combination with {\em flat} X-ray spectra whereas
NLSy1s show a rather large scatter in spectral slope and many 
are as flat as `normal' Seyfert\,1s (see particularly Xu et al. 1999).}} 
Further, there are some correlations between the optical
emission line properties in the sense that {\em small} widths of BLR lines appear 
to go hand in hand with {\em strong} FeII complexes and {\em weak} [OIII]/H$\beta$
ratios  
(e.g., Gaskell 1985, Puchnarewicz et al. 1992, Boroson \& Green 1992, 
Laor et al. 1994,
Lawrence 1997, Lawrence et al. 1997, Grupe et al. 1998).

A detailed analysis of the UV spectra of a number of NLSy1 galaxies
was carried out by Rodriguez-Pascual et al. (1997) who detected 
a broad component in the permitted UV lines (FWHM $\approxgt 5000$ km/s)
but its absence in optical lines and
favored an optically thin BLR \`{a} la Shields et al. (1995) as explanation.
They also collected IR $-$ X-ray fluxes and conclude that 
Seyferts and NLSy1s are generally very
similar concerning luminosities in different energy bands
except that NLSy1s tend to be underluminous
in the UV. 

The causes for (a) the very soft X-ray spectrum in the \ros energy band,
and for (b) the correlations among the optical emission lines and with
the X-ray properties are still under discussion. 
Whereas most of the spectral steepness in, e.g., the NLSy1 galaxy NGC\,4051 is caused 
by the presence of a warm absorber, strong soft excesses have been observed
in other sources (e.g., TON\,S180; Fink et al. 1997, Comastri et al. 1998). 
In particular,
a model that explains in detail {\em all} properties of NLSy1s within one scenario
seems to be still lacking.    
Several suggestions have been made to explain individual aspects, e.g.,
(i) a special geometry, i.e., a disk-like BLR that is viewed face-on 
(Osterbrock \& Pogge 1985, Stephens 1989, Puchnarewicz et al. 1992) or (ii) selective 
absorption of the high-velocity component of the BLR by dust (Halpern \& Oke 1987) 
to account for the small width of H$\beta$;
(iii) partial shielding of the NLR by a thick BLR (Boroson \& Green 1992)
to explain the anti-correlation of FeII and [OIII]; 
(iv) a removal or hindrance of a multi-phase BLR equilibrium by 
a steep X-ray spectrum \`{a} la Guilbert et al. (1983) (Brandt et al. 1994,
see also Komossa \& Fink 1997d) or 
(v) a scaling of BLR radius with X-ray spectral slope as in 
Wandel (1997) to explain the correlation of \G with FWHM$_{\rm H\beta}$.
BBF96 tentatively favored (vi) {\em low-mass} central black holes 
to produce a `hot' soft excess in combination with a shielded NLR
as suggested by Boroson \& Green (1992).  
Komossa \& Greiner (1995) and Komossa \& Fink (e.g., 1997a,d,e)  
studied the possibility that (vii) the steep X-ray spectra 
in the \ros band and/or the optical high-ionization iron lines 
are predominantly caused by the presence of warm absorbers,
and find, for the case of NGC\,4051, that warm absorber and coronal
line region are likely of different origin.    
Komossa \& Janek (1999) examined the influence of various EUV-X-ray spectral shapes 
on the optical emission line ratios of NLSy1s. 

\subsection {Warm absorbers}
Warm absorbers, highly ionized matter in the central region of
active galaxies (AGN), are  
an important new diagnostic tool for investigating the conditions
within the nuclei of AGN (see Fabian 1996, Komossa \& Fink 1997d for overviews).
The presence of an ionized absorber was first discovered in \ein observations
of the quasar MR 2251-178 (Halpern 1984).
With the improved spectral resolution of \ros and {\sl ASCA}, many more were
found.
They have been observed in $\sim$50\% of the well-studied Seyfert galaxies
as well as in some quasars
(e.g., Pan et al. 1990,   
Nandra \& Pounds 1992,   
Turner et al. 1993,            
Fiore et al. 1993,               
Mathur 1994,                      
Done et al. 1995,                
Cappi et al. 1996,                
Ulrich-Demoulin \& Molendi 1996, 
Komossa \& Fink 1997b,c,         
Schartel et al. 1997a).              
Signatures of ionized absorbers have also been detected in 
quite a number of NLSy1 galaxies
(e.g., Brandt et al. 1994, Pounds et al. 1994, 
Leighly et al. 1996, 1997, Guainazzi et al. 1996,
Brandt et al. 1997, Komossa \& Fink 1997a, Hayashida 1997, Iwasawa et al. 1998).

\subsection {The present study} 

Given the enigmatic properties of NLSy1s, their detailed study
is important. The sources discussed below all show some particularly
extreme behavior in terms of spectral slope or variability. 
X-ray analyses of them were either
not published previously, or with different emphasis (for details see below).

Part of the original selection criterion also was 
to check for the presence of a warm absorber (WA), 
since WAs suggest themselves as explanation for both extreme spectral
steepness in the soft X-ray band and strong spectral 
variability.{\footnote{Given 
that WAs are unambiguously detected in Seyferts,
one would naturally expect the existence of objects with {\em deeper} absorption
complexes that recover only beyond the \ros energy range
(unless an as yet unknown fine-tuning mechanism is at work,
that regulates the optical depths in the important metal ions
to a very narrow range); whether WAs are indeed present requires
a careful examination on an object-by-object basis which is presented below. 
(Note that some early arguments against WAs in \ros spectra 
simply 
mixed up some of the basic physical properties
of warm absorbers and thus led to erroneous conclusions).}  
However, we do not
only focus on this scenario. Alternatives are discussed in some detail. 
In particular, we examine the influence of different EUV-X-ray spectral
shapes on BLR multi-phase equilibrium following the suggestion of Brandt et al. (1994).

This paper is organized as follows: 
The data reduction is described in Sect. 2. In the next two
sections we present the general assumptions on which the data analysis
is based (Sect. 3) and results for the individual objects (Sect. 4).
In Sects 5.1-5.5 we give a discussion of the properties of the individual
galaxies while in Sects 5.6-5.7 consequences for NLSy1s in general are addressed. 
The concluding summary is given in Sect. 6.  
Throughout this paper, we assume $H_{0}$ = 50 km/s/Mpc and the galaxies to follow
the Hubble flow.

\section{Data reduction} 

We used all-sky survey (RASS) as well as archival and pointed serendipitous
\ros (Tr\"umper 1983) PSPC (Pfeffermann et al. 1987, Briel et al. 1994)
observations of the 
galaxies. 
The observations are summarized in Table 1.

For further analysis, the source photons were extracted
within a circular cell centered on the target source. 
The background was determined in a source-free ring around  
the target source and subtracted. 
The data were corrected for vignetting 
and dead-time, using the EXSAS software package (Zimmermann et al. 1994).
In case of RASS data the background was determined in a source-free
region along the scanning direction of the telescope.  

To carry out the spectral analysis source photons in
the amplitude channels 11-240 were binned
according to a constant signal/noise ratio of at least 3$\sigma$,
and higher for brighter sources.  

Details on the standard procedures for the reduction of \ros
data can be found in, e.g., Zimmermann et al. (1994) and Briel et al. (1994).  

   \begin{table}             
     \caption{Log of observations. $t_{\rm exp}$ gives the exposure time in ksec, $CR$ the countrate
       in cts/s. The data for NGC\,4051 are separately listed in Table 3. `S' refers to
       RASS data, `P' to PSPC observations in the pointed mode.}
     \label{obslog}
      \begin{tabular}{lllcl}
      \hline
      \noalign{\smallskip}
        object & date &  $t_{\rm exp}$ & $CR$ & obs.\\
       \noalign{\smallskip}
      \hline
      \hline
      \noalign{\smallskip}
QSO\,0117    & Dec. 28-29, 1991 & 4.5 & 0.44 & P \\
\noalign{\smallskip}
RXJ\,0134 & Dec. 1990        & 0.55& 0.30 & S  \\
\noalign{\smallskip}
       & Dec. 28, 1992 - Jan. 3, 1993 & 5.9 & 0.24 & P \\
\noalign{\smallskip}
 Mrk\,1298 & June 1-15, 1996 & 21.3 & 0.01 & P \\
\noalign{\smallskip}
 \4c & June 23-34, 1993      & 19.4 &  0.65 & P \\ 
      \noalign{\smallskip}
      \hline
      \noalign{\smallskip}
  \end{tabular}
\end{table}

  \begin{figure}[t]
\psfig{file=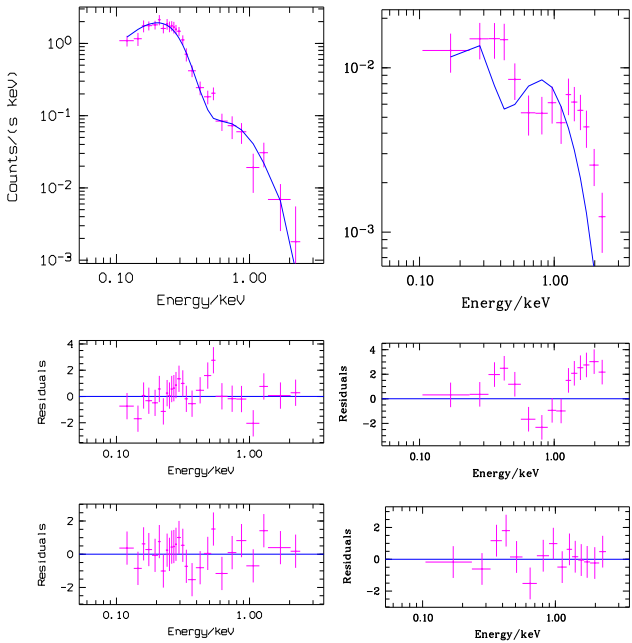,width=8.7cm}
 \caption[spec]{ X-ray spectra and residuals of the fit for \7(left) and Mrk\,1298(right). 
The upper panel gives the observed X-ray spectrum of each galaxy
(crosses) and the model fit (solid line). The lower panel shows the residuals.
\7 (left): upper panels: single powerlaw, lowest panel: warm-absorbed flat powerlaw.
Mrk\,1298 (right): upper panels: single powerlaw,
 lowest panel: warm-absorbed flat powerlaw. 
} 
\label{spec}
\end{figure}

\section {Data analysis: general assumptions}

Several models were applied to the X-ray spectra of the galaxies:
(i) a single powerlaw of the form $\Phi \propto E^{\Gamma_{\rm x}}$ 
(which, even if not the correct 
description, always provides a useful means of judging the steepness
of the spectrum), (ii) a powerlaw plus soft excess parameterized
either as black body emission or by the accretion disk model
available in EXSAS (Zimmermann et al. 1994), and (iii) a warm
absorber model. The latter was calculated with Ferland's (1993)
code {\sl Cloudy} (see Komossa \& Fink 1997a,b for details). 
The following assumptions were made:
The warm absorber is assumed to be of constant density $\log n_{\rm H}=9.5$,
of solar abundances according to Grevesse \& Anders (1989)
(if not mentioned otherwise), and to be illuminated
by the continuum of the central point-like energy source. The spectral
energy distribution from the radio to the gamma-ray region
consists of our mean AGN continuum (Komossa \& Schulz 1997)
of piecewise powerlaws with, in particular, an energy index
$\alpha$$_{\rm uv-x} = -1.4$ in the EUV and an X-ray photon index $\Gamma$$_{\rm x}$
which is either directly determined from X-ray spectral fits or fixed to --1.9.
The fit parameters of the warm absorber are its column density
$N_{\rm w}$ and the ionization parameter $U=Q/(4\pi{r}^{2}n_{\rm H}c)$.
In case of the dusty warm absorber models the dust composition and grain 
size distribution were chosen like
in the Galactic diffuse interstellar medium (Mathis et al. 1977)
as incorporated in
{\em Cloudy} (Ferland 1993), and the metal abundances
were depleted correspondingly (see Komossa \& Fink 1997b,c for details).  

\section {Data analysis results: individual objects} 

Below, we first provide a brief review of the multi-wavelength properties
of the individual sources and then report the results from our analysis
of the X-ray data. 

\subsection {QSO 0117-2837}

QSO 0117-2837 (1E 0117.2-2837) was discovered as an X-ray source by \ein and is
 at a redshift of $z$=0.347 (Stocke et al. 1991).
Grupe (1996) classified it as NLSy1.
It is serendipitously located in one of the \ros PSPC pointings; the
steep X-ray spectrum was briefly noted by Schwartz et al. (1993)
and Ciliegi \& Maccacaro (1996).
We present here the first detailed analysis of the \ros observations
of this AGN.

When the X-ray spectrum is fit by a single powerlaw
continuum with Galactic cold absorption of 
$N_{\rm Gal} = 1.65\,10^{20}$ cm$^{-2}$ (Dickey \& Lockman 1990), 
we derive a photon index \G $\simeq -3.6$
($-4.3$, if $N_{\rm H}$ is treated as free parameter).
The overall quality of the fit is good ($\chi{^{2}}_{\rm red} = 0.8$), but there are
slight systematic
residuals around the location of absorption edges.

A successful alternative description is a warm-absorbed flat powerlaw
of canonical index. 
We find a very large column density $N_{\rm w}$ in this case, and the contribution
of emission and reflection is no longer negligible; there is also some
contribution to Fe K$\alpha$. For the pure absorption model, the best-fit values
for ionization parameter and warm column density are $\log U \simeq 0.8$,
$\log N_{\rm w} \simeq 23.6$ ($N_{\rm H}$ is now consistent with the  Galactic value),
with $\chi{^{2}}_{\rm red}$ = 0.74. Including the contribution of emission
and reflection for 50\% covering of the warm material as calculated with {\em Cloudy} gives
$\log N_{\rm w} \simeq 23.8$ ($\chi{^{2}}_{\rm red}$ = 0.65).
We note that for these large column densities, the optical depth to
electron scattering becomes significant.  The main purpose of the present study
was to check under which conditions a warm absorber model fits at all;
more detailed modelling should await the availability of deeper 
observations and improved spectral resolution. 
%
\begin{table*}   
  \caption{Comparison of different spectral fits to \7, \0134 and Mrk\,1298:
  (i) single powerlaw (pl),
  (ii) accretion disk model after Shakura \& Sunyaev (1973), and (iii) warm absorber.
  \G was fixed to --1.9 in (ii) and (iii),
  except for \0134, where \G = --2.2 (see text). Instead of individual error
  bars we provide several models that successfully describe the data. 
 }
  \begin{tabular}{clllllllll}
  \noalign{\smallskip}
  \hline
  \noalign{\smallskip}
   name &  \multicolumn{3}{l}{~~~~~~~~~~powerlaw$^{(1)}$~~~~~~~~~~} &
                       \multicolumn{3}{l}{~~~~~~~~~~acc. disk + pl$^{(2)}$~~~~~~~~~~~~~~~~~} &
                         \multicolumn{3}{l}{~~~~~~warm absorber$^{(2)}$~~~~~~~~} \\
  \noalign{\smallskip}
  \noalign{\smallskip}
         & $N_{\rm H}^{(3)}$~~ & \G~~~~ & $\chi^2_{\rm red}$ & $M_{\rm BH}^{(4)}$~~ &
                                       ${\dot M}\over{{\dot M_{\rm edd}}}$~ &
                                       $\chi^2_{\rm red}$ &
                                       log $U$~~ & log $N_{\rm w}$ & $\chi^2_{\rm red}$ \\
  \noalign{\smallskip}
  \hline
  \hline
  \noalign{\smallskip}
    \7 & 0.30 & --4.3 & 0.8 & 6 & 0.6 & 0.7 & ~~0.8 & 23.6 & 0.7 \\
  \noalign{\smallskip}
  \hline
  \noalign{\smallskip}
    RX\,J0134--4258$^{(5)}$~~ & 0.16 & --4.4 & 0.5 & 1 & 0.1 & 0.6 &~~0.5 & 23.1 & 0.6 \\
  \noalign{\smallskip}
    RX\,J0134--4258$^{(6)}$~~ & 0.16 & --2.2 & 1.3 &  &  &  & &  &  \\
  \noalign{\smallskip}
  \hline
  \noalign{\smallskip}
    Mrk\,1298                 & 0.44 & --2.6 & 4.3 & 1 & 0.02 & 3.5 & --0.3 & 22.2 & 0.95  \\ 
  \noalign{\smallskip}
  \hline
  \noalign{\smallskip}
     \end{tabular}
  \label{tabn}

  \noindent{\small
  $^{(1)}$$N_{\rm H}$ free, if $>$ $N_{\rm H}^{\rm Gal}$ ~ $^{(2)}$$N_{\rm H}$
fixed to $N_{\rm H}^{\rm Gal}$ ~ $^{(3)}$in 10$^{21}$cm$^{-2}$ ~ $^{(4)}$in 10$^5$M$_{\odot}$, fixed~
$^{(5)}$survey obs. ~ $^{(6)}$pointed obs.}
\end{table*}
%
Several strong EUV emission lines are predicted to arise from the warm material.
Some of these are:
FeXXI$\lambda$2304/H$\beta_{\rm wa}$ = 10, HeII$\lambda$1640/H$\beta_{\rm wa}$ = 16,
FeXXI$\lambda$1354/H$\beta_{\rm wa}$ = 37,    
FeXVIII$\lambda$975/H$\beta_{\rm wa}$ = 16, NeVIII$\lambda$774/H$\beta_{\rm wa}$ = 9, and
FeXXII$\lambda$846/H$\beta_{\rm wa}$ = 113.
No absorption from CIV and NV is expected to show up. Both elements are more highly
ionized.

Alternatively, the spectrum can be fit with a flat powerlaw
plus soft excess (Table 2). E.g., assuming a black body shape we
derive $KT_{\rm bb} = 0.1$ keV for $N_{\rm H} = N_{\rm Gal}$ ($\chi{^{2}}_{\rm red} = 0.7$).  

An analysis of the temporal variability reveals constant source
flux within the 
1$\sigma$ error during the observation.

\subsection {\0134}

Discovered in the \ros survey (Greiner 1996), the object was optically identified
as NLSy1 galaxy (Grupe 1996) with redshift $z$=0.237.
The later pointed PSPC observation
led to the detection of strong spectral variability 
(Greiner 1996, Grupe 1996, Mannheim et al. 1996, Komossa \& Fink 1997d).
Here, we present the first detailed analysis of the X-ray properties
of this peculiar source.{\footnote{First   
results of the present study 
were reported by Komossa \& Fink (1997d),
Komossa \& Greiner (1999), and Komossa et al. (1999a). Another
study by Grupe et al. (1999b) in underway who also report the
detection of \0134 at radio wavelengths. }}  
The kind of variability of \0134 is rare, and provides
important constraints on the intrinsic X-ray emission mechanisms
and/or the properties of surrounding reprocessing material.

  \begin{figure}  
\psfig{file=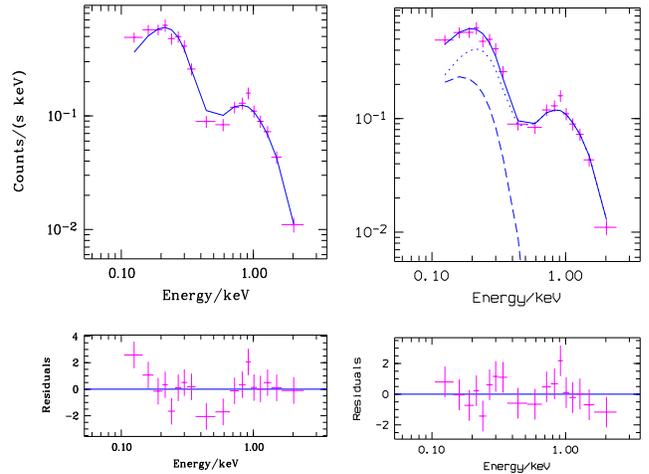,width=8.8cm}
\caption[spec_0134]{ X-ray spectrum (pointed obs.) of \0134 and residuals.
{\em Left}: The upper panel gives the observed X-ray spectrum (crosses) and powerlaw model
fit (solid line), the lower panel the residuals. {\em Right}: The same for 
a powerlaw plus black body fit (the quality of the fit is improved,
but some systematic residuals around 0.4-0.9 keV remain). The spectrum was binned to a 
signal/noise of 8$\sigma$ per bin. The amount of cold absorption was fixed to the Galactic
value.  
  }
\label{spec_0134}
\end{figure}

  \begin{figure}  
\psfig{file=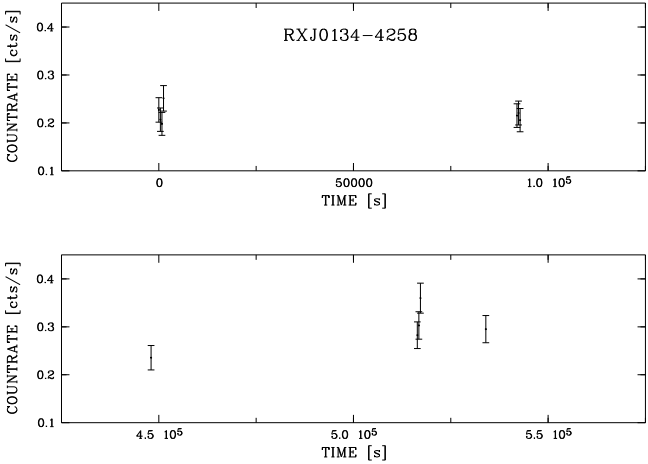,width=8.9cm}
 \caption[light_0134]{
X-ray lightcurve of \0134 (pointing) binned to time intervals of 400s.
The time is measured in seconds from the start of the observation.   }
\label{light_0134}
\end{figure}

\vskip0.2cm
\noindent{\em RASS.}
When fit by a single powerlaw, the spectrum of \0134 turns out to be one
of the steepest among NLSy1s with \G $\approx$ --4.4 (absorption was fixed
to the Galactic value in the direction of \0134, $N_{\rm Gal} = 1.59\,10^{20}$ cm$^{-2}$). 
A warm-absorbed, intrinsically {\em flat} powerlaw provides a successful
alternative fit to the RASS data. 
Due to the low number of available photons, a range of possible combinations
of $U$ and $N_{\rm w}$ explains the data
with comparable success. 
A large
column density $N_{\rm w}$ (of the order 10$^{23}$ cm$^{-2}$) is needed to
account for the ultrasoft observed spectrum.
When we fix \G = $-2.2$, the value observed during the later pointing,
and use  $N_{\rm H}$ = $N_{\rm Gal}$, we obtain $\log N_{\rm w} = 23.1$ and $\log U = 0.5$.
This model gives an excellent fit ($\chi{^{2}}_{\rm red}=0.6$).

A number of further models were compared with the observed spectrum. 
E.g., an accretion disk model was fit. Again, we fixed \G=$-2.2$. 
The black hole mass is not well constrained by the model
and was fixed
($10^{5}$ M$_{\odot}$). We find 
${\dot M}\over{{\dot M_{\rm edd}}} \simeq$ 0.1 and, again,
a very good fit is obtained (Table 2). 
If instead the spectrum is fit by a single black body, 
one derives a temperature $kT \simeq 0.07$ keV. 

\vspace*{0.3cm}

\noindent{\em Pointed observation.} 
The fit of a single powerlaw to the spectrum of \0134 
yields a photon index \G = -2.2 ($\chi^2_{\rm red}$ = 1.4), 
{\em much flatter} than during the RASS observation.
The amount of cold absorption was fixed to the Galactic 
value (if treated as free parameter, the Galactic value
is underpredicted).  
For this model fit, two kinds of residuals are visible:
(i) the first data point (below 0.15 keV) indicates a higher
countrate than predicted by the model. This data point
significantly influences the value of $\chi^2_{\rm red}$, and
if it is excluded from spectral fitting, we obtain $\chi^2_{\rm red}$ = 1.0
and \G = $-2.1$.    
Formally, a very low-temperature soft excess could be present
in the spectrum of \0134. Indeed, such a model can be fit
with $kT \simeq$ 0.1 keV.
Hints for a similar very soft excess have been found in the
\ros spectra of TON\,S180 (Fink et al. 1997) and NGC\,4051 (Komossa \& Fink 1997d).
However, since such a component is essentially only 
constrained by the first few data bins we 
do not discuss this possibility in further detail. Another possibility
is uncertainties in the calibration at these low energy channels.  
The second deviation from the powerlaw is (ii) an underprediction 
of the countrate in the energy range $\sim$0.4--0.9 keV (Fig. \ref{spec_0134})
indicative of the presence of absorption edges,
as observed in AGNs where warm absorbers are present.
However, again, the deviations from the powerlaw are only defined by
few bins, and we thus assume in the following
that the spectrum during the pointed observation essentially
represents the intrinsic, un-distorted continuum (a complete description
might invoke both, a weak soft excess and weak warm absorption, but 
fitting such models would definitely be an overinterpretation of \ros data). 

Temporal analysis:
The countrate during the pointed observation turns out to be
variable by about a factor 2.
The lightcurve is displayed in Fig. \ref{light_0134}.

  \begin{figure}[thbp]
\psfig{file=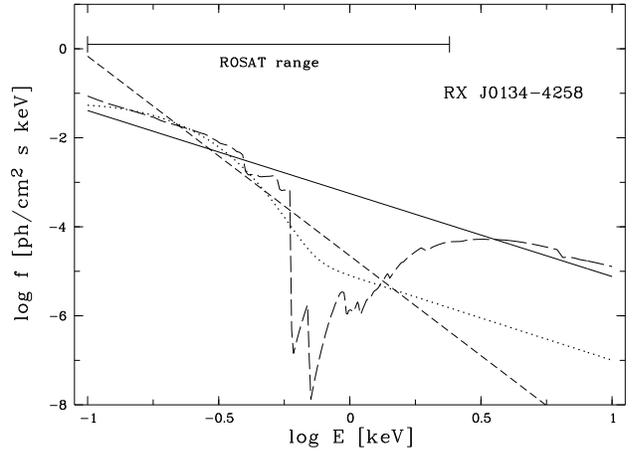,width=8.6cm}
 \caption[def_0134]{Comparison of different X-ray spectral fits 
successfully applied to the \ros survey
observation of \0134 (broken lines) and the pointed observation (solid line).
Short-dashed line: single powerlaw with $\Gamma_{\rm x} = -4.4$; long-dashed: warm-absorbed
flat powerlaw; dotted: powerlaw plus soft excess, parameterized as a black body.
All models are corrected for Galactic absorption. }
\label{def_0134}
\end{figure}

\subsection {NGC\,4051}

NGC\,4051 has been classified as Seyfert\,1.8 (e.g., Rosenblatt et al. 1992)
or NLSy1 (e.g., Malkan 1986) and is at a redshift of $z$ = 0.0023.
This galaxy has been been observed with all major X-ray satellites 
(e.g., Marshall et al. 1983, Lawrence et al. 1985, Matsuoka et al. 1990,
Mihara et al. 1994, McHardy et al. 1995, Guainazzi et al. 1996,
Komossa \& Fink 1997a; for brief summaries of these papers see 
Sect. 1 and 5 of Komossa \& Fink 1997a).
Recently, first BeppoSAX results have been presented by Guainazzi et al. (1998a),
who report the detection of a strong drop in source flux which lasted
the whole observing interval of $\sim$2 d.

Here, we present an analysis of all \ros PSPC data of this
source, including previously unpublished observations and a homogeneous re-analysis
of published ones (McHardy et al. 1995).
Since NGC\,4051 is strongly variable in X-rays, the large set of \ros data
is very valuable to create a long-term lightcurve of this source
and to study variability mechanisms. It also provides an excellent data
base to study long-term spectral changes due to the presence of the
warm absorber and places tight constraints on the ionization state
of the warm material.

To investigate the long-term trend in the variability of NGC\,4051, in countrate
as well as in ionization parameter $U$ and column density $N_{\rm w}$
of the warm absorber, we 
have fit our warm absorber model to the individual data sets. 
We find that in the long term all features are variable,
except for the cold absorption which is
always consistent with the Galactic value within the error bars.
Ionization parameter $U$ and column density 
$N_{\rm w}$ change by about a factor of 2. The slope of the
powerlaw remains rather steep (Table 3). 

The long-term lightcurve reveals 
large-amplitude variability by a factor $\sim$30 in countrate
within the total observing interval. 
The X-ray lightcurve is displayed in Fig. \ref{light_4051}.

 \begin{figure*}[ht]
\psfig{file=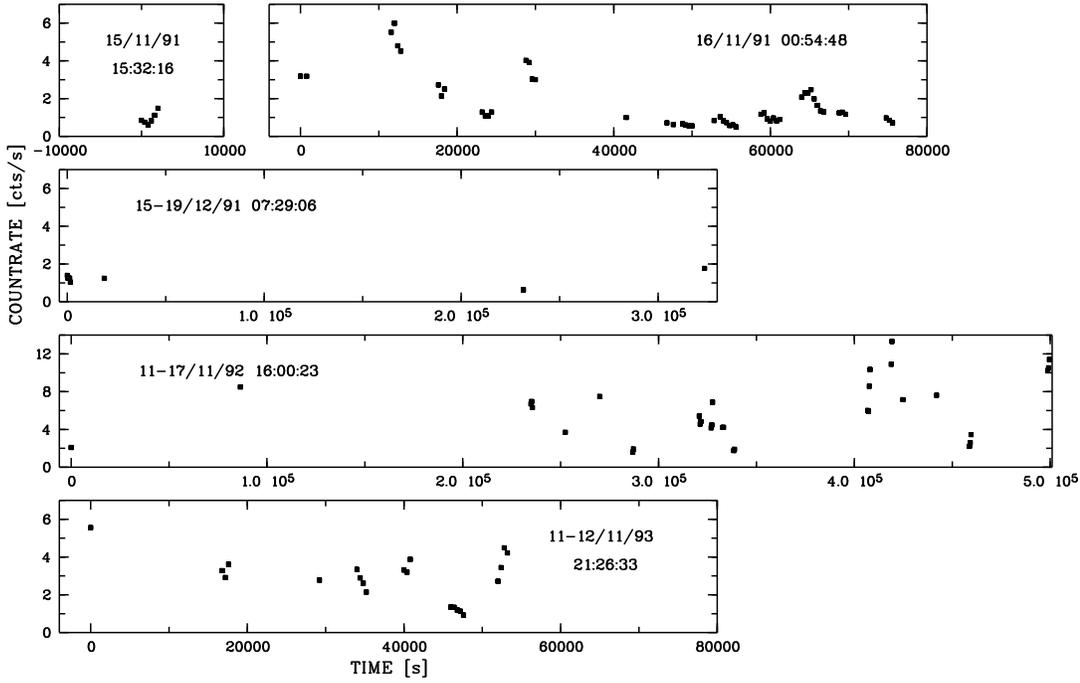,width=15cm}
 \caption[light]{Long-term X-ray lightcurve of NGC\,4051, based on all
pointed \ros PSPC observations of this source.  NGC\,4051 
is variable by a factor $\sim$30 in countrate.
The lightcurve of Nov.\,16,\,1991 was earlier
shown in McHardy et al. (1995), the one of Nov.\,1993 in
Komossa \& Fink (1997a). The time is measured in s from the beginnings of the individual
observations; the insets in each panel give the starting times. }
\label{light_4051}
\end{figure*}

   \begin{table*} 
  \caption{Log of \ros PSPC observations of NGC\,4051 and warm absorber fit results.
                  $t_{\rm exp}$ = effective exposure time, $CR$ = mean count rate,
                  $L_{\rm x}$ =  mean (0.1--2.4 keV) luminosity corrected
                  for cold and warm absorption.}
      \begin{tabular}{ccccccc}
      \noalign{\smallskip}
      \hline
      \noalign{\smallskip}
       date & $t_{\rm exp}$ [s] & $CR$ [cts/s]& log $U$ & log $N_{\rm w}$ &
              $\Gamma_{\rm x}$ & $L_{\rm x}$ [10$^{41}$ erg/s] \\
      \noalign{\smallskip}
      \hline
      \hline
      \noalign{\smallskip}
 Nov. 15, 1991 & ~2705 & 1.0 & 0.2 & 22.52 & -2.2 & 3.5\\
 Nov. 16, 1991 & 28727 & 1.6 & 0.2 & 22.45 & -2.2 & 5.4 \\
 Dec. 15 - 19, 1991 & ~4783 & 1.0 & 0.1 & 22.35 & -2.2 & 3.7 \\
 ~Nov. 11 - 17, 1992$^{*}$ & 20815 & 6.0 & 0.2 & 22.35 & -2.2 & 19.5~~\\
 Nov. 11 - 12, 1993 & 12261 & 2.9 & 0.4 & 22.67 & -2.3 & 9.5\\
      \noalign{\smallskip}
      \hline
      \noalign{\smallskip}
\end{tabular}

\noindent {\scriptsize
           $^{*}$ results of model fits uncertain due to off-axis location 
           of source}
   \end{table*}

\subsection {Mrk\,1298} 

Mrk\,1298 (PG 1126-041) is a luminous Seyfert\,1 galaxy at redshift $z$ = 0.06
(Osterbrock \& Dahari 1983).
Its optical spectrum (Rafanelli \& Bonoli 1984, Miller et al. 1992)
is characterized by strong FeII emission line complexes. 
Mrk\,1298 was part of several studies of correlations between 
strength of FeII and other spectral properties (Boroson \& Green 1992,
Wang et al. 1996 (WBB96 hereafter)). 
A UV spectrum of Mrk\,1298 was presented by Wang et al. (1999).
The \ros PSPC X-ray spectrum was first analyzed by WBB96 
who detected an absorption edge which they interpreted as arising
from a warm absorber.
We present here a more detailed analysis of the properties
of the warm absorber (see also Komossa \& Fink 1997d,e), 
including predictions of non-X-ray emission
lines expected to arise from the warm material,
and test for the presence of a {\em dusty} warm absorber.
We also analyze the temporal behavior of the X-ray flux.

A single powerlaw does not provide a successful X-ray spectral
fit. We find $\chi{^{2}}_{\rm red}$=3.3 and the amount of
cold absorption underpredicts the Galactic value in the direction 
of Mrk\,1298, $N_{\rm H}^{\rm Gal} = 4.44\,10^{20}$ cm$^{-2}$. 
If Galactic absorption is enforced, the quality of the fit 
becomes worse ($\chi{^{2}}_{\rm red}$=4.3). 
Therefore, a number of further spectral models was fit,
including an intrinsically flat powerlaw plus black-body
like soft excess. The latter model gives $\chi{^{2}}_{\rm red}$=3.4 (Table 4),
still unacceptable.

On the other hand, a warm absorber fits the X-ray spectrum well.
We fixed the photon index of the intrinsic powerlaw to \G=$-1.9$.
In a first step, cold absorption was fixed to the Galactic value. 
We then obtain $\log U \simeq -0.3$ and $\log N_{\rm w} \simeq 22.2$ 
and the quality of the fit is acceptable ($\chi{^{2}}_{\rm red}$=0.95). 
Slight systematic residuals remain at very low energies. Thus,
in a second step, \NH was treated as free parameter. 
In this case we find some excess absorption, the fit is further improved
($\chi{^{2}}_{\rm red}$=0.78), and the residuals disappear (Fig. \ref{spec}).  
The warm absorber parameters change 
to $\log U \approx -0.1$ and $\log N_{\rm w} \approx$ 22.5.
The cold absorption amounts to \NH $\simeq 0.6 \times 10^{21}$ cm$^{-2}$.

Finally we note that the model of a {\em dusty} warm absorber does not give 
a successful X-ray spectral fit provided the intrinsic powerlaw spectrum
is close to \G =$-1.9$.  

The X-ray temporal analysis (Fig. \ref{light_1298}) reveals rapid variability of the source 
with repeated changes in countrate by a factor $\sim$2 within 800 s. 

  \begin{figure}[thbp]
\begin{center}
\psfig{file=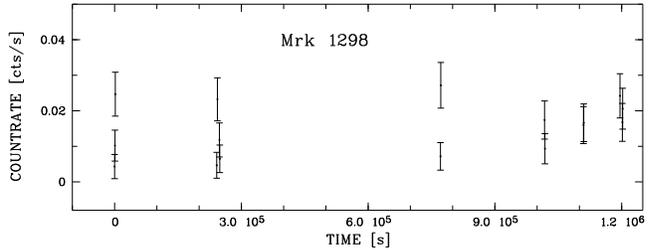,width=8.8cm}
 \caption[light_1298]{X-ray lightcurve of Mrk\,1298. Each point encloses
a time interval of 800s. Rapid variability on short timescales is
revealed.} 
\label{light_1298}
\end{center}
\end{figure}

\subsection {\4c}
\4c is a radio-loud quasar at $z$=0.104 (Riley et al. 1988). The RASS data of
this source were analyzed by Schartel et al. (1996a) who derived \G $\simeq -1.3$.
In a study of \ros and \asca data,
Brinkmann et al. (1998) confirmed the unusually flat soft X-ray \ros spectrum
(\G $\simeq -1.3~{\rm to} -1.6$; as compared to \G $\simeq -2.2$ typically seen in
nearby radio-loud quasars), found a steeper \asca powerlaw spectrum, and evidence for the
presence of a warm absorber.
We re-analyzed this source, since the description given in Brinkmann et al. 
was highly suggestive of the presence of a {\em dusty} warm absorber.

The fit of a single powerlaw model yields an acceptable fit 
($\chi{^{2}}_{\rm red} = 1.0$) but an extremely flat spectrum
with \G=$-1.4$.   
Applying the model of a {\em dusty} warm absorber to the \ros spectrum
we get a successful spectral fit with a steeper intrinsic powerlaw. 
In particular, we fixed the photon index to \G=$-2.2$ since we wanted
to test whether the data are consistent with the general
expectation for radio quasars. 
In this case we obtain a column density of the dusty warm gas
of $\log N_{\rm w}$ = 21.6 and an ionization parameter of $\log U = -0.1$
($\chi{^{2}}_{\rm red}$ = 1.0).

\section{Discussion}

We first discuss the X-ray properties of the individual objects,
in the context of other object-specific observations. 
Then, a more general discussion on NLSy1s is given.  

\subsection{ \7} 
Three models were found to fit the \ros X-ray spectrum of 
\7 successfully. A single steep powerlaw with \G $\simeq -4.3$,
a flat powerlaw plus soft excess, and a warm-absorbed
flat powerlaw. In the latter case a rather large column density
of the warm absorber is inferred, $\log N_{\rm w} \approx 23.6-23.8$.
This would make the ionized absorber in \7 the one with the
largest column density known, and suggests that other 
spectral components contribute to, or dominate, its X-ray spectral
steepness. Given the reported relations between X-ray and
UV absorption, we note that we do not predict any
UV absorption from NV and CIV for our best-fit warm absorber
model. Both elements are more highly ionized; i.e.,
the absence of UV absorption alone could not be used as
argument against a warm absorber.    

Given the very steep rise towards the blue of QSO\,0117-2837's optical
spectrum (Grupe et al. 1999a; Fig. 1a of Komossa et al. 2000),
it is tempting to speculate that a giant soft-excess
dominates the optical-to-X-ray spectrum. We strongly caution,
though, that simultaneous optical-X-ray variability studies
in other Seyferts and NLSy1s (e.g., Done et al. 1995) 
do {\em not} favor a direct relation between optical and X-ray
spectral components. Furthermore, such a giant optical-to-X-ray bump
in QSO\,0117-2837 and NLSy1s in general, would be inconsistent with the
finding of Rodriguez-Pascual et al. (1997) that NLSy1s tend to be underluminous
in the UV. 

 A most interesting peculiarity of QSO\,0117-2837 is revealed,
 when combining its X-ray and optical properties:
 Whereas its X-ray spectrum is among the steepest observed
 among NLSy1s, its H$\beta$ emission line is fairly broad.
 Fitting this line with Gaussian components it  
 is best described by a two component Gaussian with a narrow
 component of similar width as [OIII] plus a broad component
 of FWHM$_{H\beta}$ $\approx$ 4000 km/s (Komossa et al. 2000). 
 This combination of \G and FWHM$_{H\beta}$ places QSO\,0117-2837
 in a region in the popular \G versus FWHM$_{H\beta}$ diagram (e.g., Fig. 8 of BBF96)  
 which is barely populated by objects, therefore occasionally 
 referred to as `zone of avoidance'.  
  
 Its peculiar properties make QSO\,0117-2837
 a good target for follow-up 
 X-ray spectral observations with {\sl XMM} and {\sl AXAF},
 as well as for high-resolution optical observations of the
 H$\beta$ complex.

Some NLy1s have been reported to show rapid X-ray variability.
\7 shows constant source flux during the observation, though.

\subsection{ \0134}

\subsubsection{Comparison with similar objects} 

The drastic spectral variability of \0134 is rather peculiar.
Often, in AGN, the X-ray flux is variable but the 
spectral slope remains constant. 
Cases were a strong change in hard X-ray spectral shape was reported, are 
IRAS\,13224 (Otani et al. 1996), NGC\,4051 (Guainazzi
et al. 1996), and Mrk\,766 (Leighly et al. 1996).
However, in all cases the spectral index varied mainly between
\G $\simeq -2.0$ and {\em flatter} values. Further, these
sources changed countrate when changing spectral shape.  
Recently, Guainazzi et al. (1998b) presented 
observations of the Seyfert galaxy 1H0419-577 which underwent a
spectral transition from steep (\G=$-2.5$) to flat (\G=$-1.6$) between 
a \ros and a {\sl SAX} observation. 
The most similar case to \0134 we are aware of
is the observation reported by Fink et al. (1997) 
who detected changes in spectral index of the
NL quasar TON\,S180 that were not accompanied by a noticeable
change of the total soft X-ray emission.

\subsubsection{Variability mechanisms}

\paragraph{Warm absorption.}

One natural mechanism to produce the spectral variability in \0134 
is warm absorption because this is an efficient method to
produce steep X-ray spectra (e.g., by a change in ionization state
of the warm absorber). Note, that Grupe (1996) argued against
the presence of the warm absorber based on the erroneous statement
that a warm absorber could not produce a steep soft X-ray spectrum. 

Examination whether, and under which conditions, a warm absorber is
indeed a viable description of the
X-ray spectrum, and whether it is the only one,
has to be based upon detailed modeling and careful consideration of
alternatives. 

Our modelling (Sect. 4.2) leads to the following results: 
The ultra-soft state is well fit by a warm absorber with
column density $ \log N \simeq 23$. This is a factor of
about 2-3 larger than that of the well-studied warm absorber in the NLSy1
galaxy NGC\,4051 (e.g., Pounds et al. 1994, Komossa \& Fink 1997a,
and our Table 3).  

The most suggestive scenario within the framework of warm absorbers
then 
is a change in the {\em ionization state} of ionized material along the line of sight,
caused by {\em varying irradiation} by a central ionizing source.
One problem arises immediately, though: 
In the simplest case, lower intrinsic luminosity would be expected, to cause the deeper
observed absorption, in 1990. However, 
the source is somewhat brighter in the RASS observation.
(Fig. \ref{def_0134}).  Some variability seems to be usual, 
though. The countrate changes by
about a factor of 2 during the pointed observation 
(Fig. \ref{light_0134}). If one wishes to keep this
scenario, one would have to assume that the ionization state of the absorber still reflects
a preceding (unobserved) low-state in intrinsic flux.

Alternatively, and more likely, gas heated by the central continuum source
may have {\em crossed the line of sight},
producing the steep RASS spectrum, and has (nearly) disappeared in the 1992 observation.
This scenario explains most naturally the nearly constant countrate
from RASS to pointed observation, because the countrate is dominated
by the soft energy part of the spectrum (below 0.7 keV) which is
essentially unaffected by warm absorption.   
The transient passage of a BLR cloudlet would be consistent with the 
scenario proposed by Rodriguez-Pascual et al. (1997) who suggested a matter-bounded 
BLR in NLSy1s on the basis of emission line profiles and strengths.  
They derive a lower column density for the BLR clouds but since our 
best-fit X-ray warm absorber also has a higher ionization parameter, 
the hydrogen Str\"omgren sphere is shifted further out and thus the clouds
remain matter-bounded.{\footnote{It is important to note that both
these scenarios are consistent with the recent \asca observation 
of Grupe et al. (1999b) who do not detect strong absorption edges
in the \asca spectrum of \0134 (the upper limits they report are {\em not} very tight,
though): In case of a non-equilibrium warm absorber the actual
depths of absorption edges depend on the unknown flux history; in
the more likely case of a transient cloud passage the absorber has simply 
{\em left our line-of-sight}.}}

\paragraph{Alternatives:}                          

The short duration of the RASS observation has to be kept in mind,
and both, an intrinsically steep powerlaw and a strong soft excess fit the
X-ray spectrum as well. 
Variability in only one component seems to be problematic, though, since the 
nearly constant countrate has to be accounted for. 

A spectral change with constant countrate is reminiscent of one class 
of Galactic black hole transients (the one in which both spectral
components change
simultaneously as to mimic constant countrate; e.g., Tanaka 1997).
In fact, the potential similarity of NLSy1s with Galactic black hole
candidates has been repeatedly mentioned (starting with Pounds et al. 1995),
but has never been explored in more detail. 
We do not follow this one further, since the analogy between 
NLSy1s and Galactic black hole candidates 
does not seem to go very far (e.g., p.\,411 of Brandt \& Boller 1999). 

Finally, it is also possible that the constant countrate is pure
coincidence: Both, variable soft excesses (see, e.g., 1E1615+061 for
an extreme example) {\em and} variable powerlaws (often of constant shape)
have been observed in AGN and these two might have compensated
each other to produce nearly constant total countrate (this
seems to be the model favored by Grupe 1996; their Fig. 8.11).

\subsection{ NGC\,4051}

The variability amplitude of NGC\,4051, a factor 30 over the measured
time interval, is fairly large. 
During all individual \ros observations, the source is
variable. No long-term very low state as recently reported
by Guainazzi et al. (1998a) occurred. 

The warm absorber properties, averaged over individual
observations, are found to be quite constant, with
changes less than about a factor of two in column density
and ionization parameter. 
This is consistent with the finding of Komossa \& Fink (1997a)
that the bulk of the warm material in NCC\,4051 does not react to 
short-timescale changes in the ionizing luminosity
and thus the bulk of the ionized absorber must be of low density,
or alternatively, the warm material is out of photoionization equilibrium.  
The latter possibility was also suggested by Nicastro et al. (1999) 
based on shorter-timescale variability behavior of NGC\,4051.
Recently, Contini \& Viegas (1999) presented detailed
multi-wavelength modelling of NGC\,4051, including
in their models the presence of shocks besides photoionization.

\subsection{ Mrk\,1298}

It is interesting to note that Mrk\,1298 exhibits all characteristics of
a NLSy1 galaxy except that its observed FWHM of H$\beta$, 2200 km/s, just escapes
the criterion of Goodrich (1989).

A warm absorber fits well the X-ray spectrum of this galaxy,
whereas a powerlaw plus black-body-like soft excess does not.
This also holds for
further possible shapes of the soft excess. 

In the following, we give some predictions
made by the warm absorber scenario, in terms of line emission
and absorption. 
Depending on the covering factor of the warm absorber, 
the ionized material might contribute to high-ionization
emission lines in the optical--EUV spectral region (e.g.,
Komossa \& Fink 1997a,d,e). 
Among the strongest predicted lines are 
[FeXIV]$\lambda$5303/H$\beta_{\rm wa}$ = 3, 
(NV$\lambda$1240+FeXII$\lambda$)/H$\beta_{\rm wa}$ = 13
and OVI$\lambda$1035/H$\beta_{\rm wa}$ = 284.
However, the warm absorber is matter bounded and the
total emissivity in H$\beta$ is fairly small when compared
to the observed H$\beta$-luminosity of 
$L_{\rm H\beta}^{\rm{obs}} \simeq 10^{42.86}$ erg/s (Rafanelli \& Bonoli 1984;
the value given in Miller et al. 1992 is a factor 1.6 lower). 
Scaled to observed H$\beta$, the strongest predicted line is 
OVI$_{\rm{wa}}$/H$\beta_{\rm{obs}} \approx$ 0.4.

WBB96 mention the presence of UV absorption lines 
in an IUE spectrum of Mrk\,1298.  
The following equivalent widths of UV absorption lines are predicted for the 
best-fit warm absorber model (see also the discussion in Wang et al. 1999): 
$\log W_{\lambda}$/$\lambda \simeq -2.9$
in CIV and Ly$\alpha$ and $\log W_{\lambda}$/$\lambda \simeq -3.0$ in NV
(adopting a velocity parameter $b$ = 60 km/s).  
This is assuming all spectral steepness is indeed caused by the warm absorber.
If an additional soft excess is present (note that just a powerlaw plus black-body-like
soft excess does {\em not} fit the \ros spectrum; but in case more than two spectral
components are allowed, fits are not well constrained due to the limited PSPC
spectral resolution) the contribution from the warm absorber would be less,
since a lower column density would be inferred from spectral fits.

\subsection{ \4c}

There is growing evidence that several ({\em but not all}) warm absorbers contain dust.
The reported
individual cases are IRAS\,13349+2438 (Brandt et al. 1996, 
Komossa \& Greiner 1999, Komossa et al. 1999b),
NGC\,3227 (Komossa \& Fink 1997b, George et al. 1998), NGC\,3786 (Komossa \& Fink 1997c),
MCG\,6-30-15 (Reynolds et al. 1997), IRAS\,17020+4544 (Leighly et al. 1997,
Komossa \& Bade 1998).

The advantage of invoking dust mixed with the warm absorber
in \4c is the steeper 
intrinsic 
powerlaw spectrum (\G $\simeq -2.2$) required to compensate for the `flattening effect' 
(Komossa \& Fink 1997b)  
of dust, as compared to a single powerlaw fit which gives a peculiarly flat spectrum
(\G $\simeq -1.4$).  

A steep intrinsic spectrum has the advantage of being close 
to the \asca hard-energy value of \G derived for this source (Brinkmann et al. 1998)
and the general expectations for
nearby radio-loud quasars in the \ros band 
(\G $\simeq -2.2$; e.g., Schartel et al. 1996a, Brinkmann et al. 1997, Yuan 1998).
Better spectral resolution soft X-ray data are
needed to distinguish between both possibilities, an intrinsically
flat powerlaw, or a dusty warm absorber.

\subsection{X-ray spectral complexity in NLSy1 galaxies} 

Based on limited spectral resolution in the soft X-ray band, 
early models attempted to explain the X-ray spectral steepness of NLSy1s 
with one component only; either (i) a single steep powerlaw,
or (ii) a strong soft excess on top of a flat powerlaw (e.g., Puchnarewicz
et al. 1992, BBF96), 
in analogy to Seyferts (e.g., Walter et al. 1994) and quasars (e.g., Schartel et al. 1996a,b, 1997b)
which were believed to have soft X-ray excesses, or (iii) heavily
warm-absorbed flat powerlaws (e.g., Komossa \& Greiner 1995, Komossa \& Fink 1997a,d,e).
One of the comfortable properties of both,  the soft excess plus flat powerlaw 
and the warm-absorbed flat powerlaw interpretation, is the presence of enough 
X-ray photons to account for the strong observed
FeII emission in NLSy1s.   
It does not immediately explain the occasionally observed trend of {\em stronger} FeII
in objects with {\em steeper} X-ray spectra, but it is interesting to note
that Wang et al. (1996) find a trend for stronger FeII to preferentially occur in objects
whose X-ray spectra show the presence of absorption edges.

However, there were early indications of spectral complexity of NLSy1s
(e.g., Brandt et al. 1994, Komossa \& Fink 1997a,d; see also Vaughan et al. 1999). 
E.g., a detailed study of NGC\,4051, a bright source for which several 
deep \ros observations were performed, revealed all three of the spectral
components to be simultaneously present: The spectral steepness is dominated by the warm absorber,
but the index of the underlying powerlaw can become as steep as 
\G $\approx -2.3${\footnote{Given the recent reports of occasional discrepancies
between the powerlaw indices derived from fitting \ros and \asca data,
in the sense that \ros spectra tend to be steeper, this might also
affect NGC\,4051's spectrum. We note, however, that during the Nov. 1992
\ros observation of this galaxy, the powerlaw was in its steepest observed
state even when comparing with other \ros observations of this source.}}  
and an additional soft excess ist present in source high-states (Pounds et al. 1994,
Komossa \& Fink 1997a; the latter authors additionally provided evidence 
for an EUV bump component based on photon counting arguments).   
Further complications concerning the X-ray spectra of NLSy1s
have emerged recently, via the suggestion
of {\em dusty} warm absorbers in some NLSy1-like galaxies (Brandt et al. 1996, Leighly et al. 1997,
Komossa \& Bade 1998).

We have examined two further NLSy1s observed with {\sl ROSAT}, 1ZwI and 
PHL\,1092, and find that neither a flat powerlaw with soft excess
nor a warm absorber can account for {\em most} of the spectral steepness.
The sources are best described by a single steep spectral component. 

As suggested by Komossa (1997) these spectral components of NLSy1s
may be linked in the sense that a more polar view on the accretion 
disk (e.g., Fig. 3 of Madau 1988) causes the soft excess component to be more
pronounced, while along the funnels of the disk outflows are driven
which cause the characteristic absorption edges of warm absorbers
if viewed along the line-of-sight against the continuum source. 

Except for NGC\,4051 the faintness of the objects of 
the present study (and many other \ros observed 
ultrasoft sources) does not allow to perform n-component spectral fits.
However, the one-component models presented here still provide 
upper limits on the contribution of each single component (steepness of powerlaw,
strength of soft excess, column density of warm absorber). 

Given the few cases that have been observed with high X-ray spectral resolution 
and sufficient countrates,  other approaches to distinguish between
different EUV-X-ray spectral shapes are important. 
Such an approach is described in the following section.

 \begin{figure}[h]
\hspace*{0.5cm}
\psfig{file=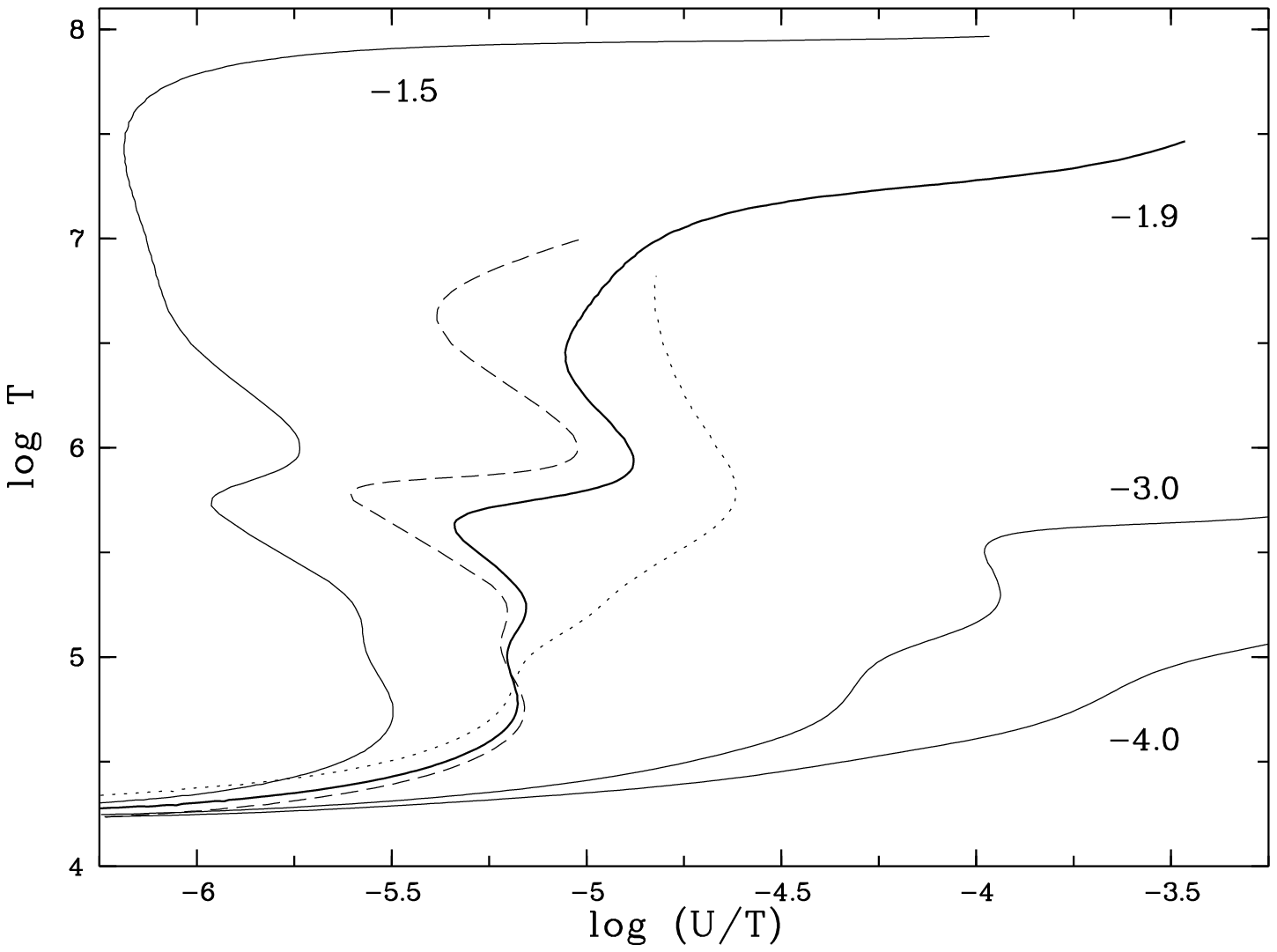,width=7.8cm}
\caption[equ_a] {Gas equilibrium curves. The X-ray spectral
shape and the gas metal abundances were varied.
The solid lines correspond to mean Seyfert continua with energy index
$\alpha_{\rm uv-x}$=--1.4, varying photon index $\Gamma_{\rm x}$ as indicated in the figure
and solar chemical abundances.
Dotted line: metal abundances of $Z$=0.3$\times$$Z_\odot$; dashed line:
$Z$=3$\times$$Z_\odot$. }
\label{equ_a}
\vspace*{0.4cm}
\hspace*{0.5cm}
\psfig{file=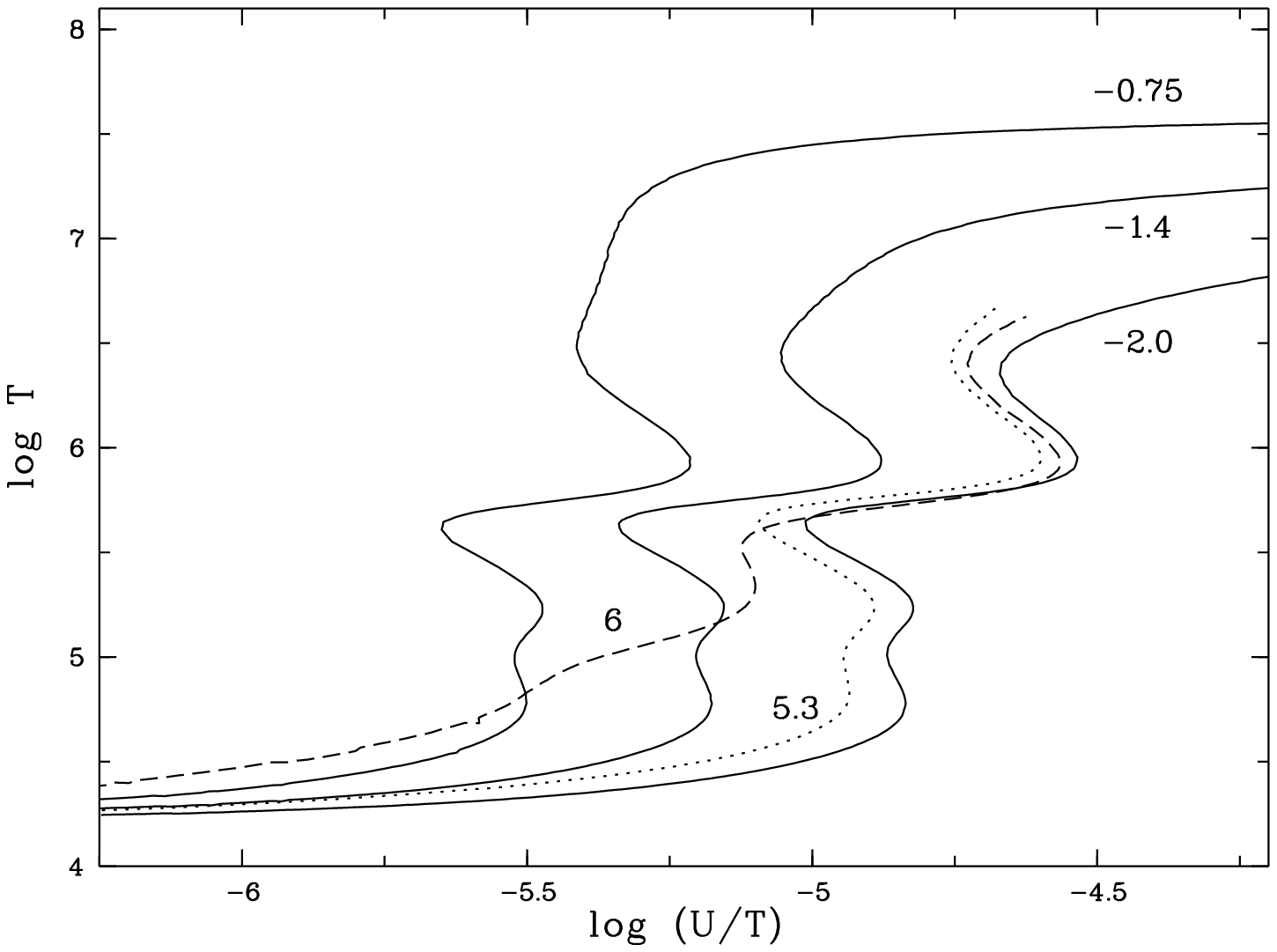,width=7.8cm}
\caption[equ_b]{ Equilibrium curves for various EUV shapes of the ionizing
spectrum illuminating the gas clouds. 
Solid lines: Continua with photon index $\Gamma_{\rm x}$ = --1.9 and
varying $\alpha_{\rm uv-x} = -0.75,-1.4,-2.0$ as given in the figure. 
Broken lines: Black-body added
to a mean Seyfert continuum with a 50\% contribution to the total luminosity and
a temperature of $10^6$K (dashed line) and 2\,$10^5$K (dotted line). 
Solar abundances were assumed.
As can be seen, steep X-ray spectra remove the multi-valued behavior of the
curves, and thus the possibility of multiple phases in pressure balance.}
\label{equ_b}
\end{figure}

\subsection {EUV -- soft X-ray spectral shape and stability of broad line clouds}

One suggestion to link the apparently steep soft X-ray spectra with the small FWHM 
of the broad lines in NLSy1s was the influence of the X-ray spectral
shape on multi-phase BLR cloud equilibrium,
especially the hindrance of BLR formation due to illumination 
by a steep X-ray spectrum 
(Brandt et al. 1994).   
Here, we test this suggestion based on calculations
carried out with the code {\em Cloudy}. 
In particular, we investigate how different EUV-X-ray spectral shapes 
change the range in which a multi-phase equilibrium is possible
and attempt to distinguish, within the limits of this scenario,
between different suggested spectral models. 

The thermal stability of broad line clouds can be examined 
by studying the behavior of temperature $T$ in dependence
of pressure, i.e., $U/T$ (e.g. Guilbert et al. 1983, their Fig. 1;
for a general discussion see also Krolik et al. 1981, Netzer 1990, Reynolds \& Fabian 1995,
Komossa \& Fink 1997a,d).
In case $T$ is multi-valued for constant $U/T$, and the gradient of
the equilibrium curve is positive, several phases may exist in
pressure balance.

To test this idea, we have calculated equilibrium curves
for the BLR gas for intrinsically steep X-ray spectra or spectra with a black-body-like
soft excess. 
Additionally, the metal abundances were varied, which
affect the cooling of the gas.
Results are shown in Figs. \ref{equ_a}, \ref{equ_b}.
The parts of the equilibrium curve with negative gradient
correspond to thermally unstable equilibria. 
It is well known that the original Krolik-McKee-Tarter models face problems when a
more realistic continuum shape is used, in the sense that a pressure balance between a cold,
photoionization heated, and a hot, Compton heated phase no longer exists (see Fig. \ref{equ_a},
e.g. $\Gamma_{\rm{x}}$ = --1.9 compared to $\Gamma_{\rm{x}}$ = --1.5).
Reynolds \& Fabian (1995) point to the existence of an intermediate-temperature stable region
where $U/T$ is multi-valued. We find that this intermediate region disappears for
steep X-ray spectra.{\footnote {We only explore the rough trends here. 
The detailed shape of the equilibrium curve in the intermediate-temperature
region also depends on the details of the heating-cooling processes and 
reflects to some extent the completeness with which these are implemented in the
code used (see Ferland et al. 1998, Kingdon \& Ferland 1999).}} 
 Due to the relatively weaker X-ray flux and
the fact, that the value of $U$ is dominated by the EUV flux near the Lyman limit,
the gas remains longer in the `photoionization-heating, collisional-deexcitation-cooling' phase.
The same holds for a continuum with a hot soft X-ray excess (Fig. \ref{equ_b}).  

These studies reveal that several spectral
shapes and gas metal abundances lead to similar results,
but show that the mechanism suggested by Brandt et al. (1994) works in general. 
More detailed models should then await better knowledge of the 0.1-10 keV
X-ray spectral shape.

\section {Summary and conclusions}

We have presented a study of the X-ray spectral and temporal 
properties of four NLSy1-like 
galaxies and a radio-loud quasar.
The results can be summarized as follows: 

{\em \underline {QSO\,0117-2837}}. 
This NLSy1 galaxy shows a very steep X-ray spectrum
with \G $\simeq -4.3$ when fit with a simple powerlaw,
among the steepest spectra reported for NLSy1s. 
Alternatively, the spectrum
can be described by a flat powerlaw with 
soft excess of $kT_{\rm bb} \simeq 0.1$ keV,
or a warm-absorbed flat powerlaw. 
The extreme X-ray spectral steepness of QSO\,0117-2837 is
unexpected in the light of its fairly large FWHM of H$\beta$.
With these properties QSO\,0117-2837 fills the `zone of avoidance'
in the FWHM$_{H\beta}$--\G diagram. 
The X-ray lightcurve of \7 reveals constant source flux. 

{\em \underline {RX\,J0134-4258}}. This source underwent a drastic X-ray spectral
transition from steep  
(\G $\simeq -4.4$) to flat (\G $\simeq -2.2$) between
two \ros observations separated by 2 yr, while 
the mean countrate remained nearly constant. We    
examined several scenarios that might account for
this peculiar behavior, with focus on the presence of a warm
absorber. We find that a reaction
of the ionized material to continuum changes requires 
non-equilibrium effects to be at work.
Alternatively, and more likely,
a cloud of warm gas may have passed our line of sight. 
This latter scenario shares some similarity with the one proposed by Rodriguez-Pascual
et al. (1997) based on UV observations of a sample of NLSy1s.  
Variability of both components in the framework of a 
powerlaw-plus-soft-excess spectral description provides an alternative explanation.  

{\em \underline {NGC\,4051}}. We analyzed all \ros PSPC observations of this well-known Seyfert galaxy
including previously unpublished ones. Variability by a factor of $\sim$30 in 
countrate is detected. The mean X-ray luminosity varies less but still by a factor of $\sim$7,
whereas the properties of the warm absorber ($U$, $N_{\rm w}$; averaged over individual pointings)
change by only a factor of $\sim$2, indicating that the bulk of the 
warm absorber is either of low density or out of photoionization equilibrium.    

{\em \underline {Mrk\,1298}}. The \ros PSPC spectrum of this source shows clear signs
of warm absorption, confirming Wang et al. (1996). 
Based on detailed warm-absorber modelling, we predict the 
contribution of the ionized absorber to opt-UV-EUV emission lines which is 
found to be quite weak. 
Repeated rapid variability by a factor $\sim$2 within time intervals of 800s is detected.  

{\em \underline {4C\,+74.26}}. We considered this quasar 
as a candidate for a {\em dusty} warm absorber.
We show that such a model successfully fits the \ros X-ray spectrum of this source
and resolves the discrepancy between the underlying powerlaw index derived from
\ros and \asca observations. 

{\em \underline {X-ray spectral complexity in NLSy1s and multi-phase }}
{\em \underline {BLR equilibrium}}.
Given the increasing spectral complexity of NLSy1 galaxies, with often
two or even three components contributing to the X-ray spectral shape,
we examined the influence of different spectral shapes on BLR cloud
multi-phase equilibrium, in an attempt to determine which spectral shapes 
dominates on average (within the limits of this scenario).
We find that both, a steep powerlaw spectrum, and a strong EUV-X-ray
excess component narrow down the range where a stable multi-phase 
equilibrium is possible.  

\begin{acknowledgements}
I, St.\,K., gratefully remember Henner Fink for
introducing me to the work with X-ray data, for many discussions
and helpful advice.
Henner Fink passed away in December 1996.
We thank Hartmut Schulz and Dirk Grupe  
for a critical
reading of the manuscript,  Dirk Grupe for providing a preprint
of his paper on RXJ0134 prior to publication,  
Gary Ferland for providing {\em Cloudy},
and Emmi Meyer-Hofmeister for her continuing kind interest.
The \ros project has been supported by the German Bundes\-mini\-ste\-rium
f\"ur Bildung, Wissenschaft, Forschung und Technologie 
(BMBF/DLR) and the Max-Planck-Society.
This research has made use of the NASA/IPAC extragalactic database (NED)
which is operated by the Jet Propulsion Laboratory, Caltech,
under contract with the National Aeronautics and Space
Administration. \\
Preprints of this and related papers can be retrieved from our webpage 
at http://www.xray.mpe.mpg.de/$\sim$skomossa/
\end{acknowledgements}


\begin{thebibliography}{}

\bibitem{} Bade N., Fink H.H., Engels D., et al., 1995, A\&AS 110, 469

\bibitem{} Becker C.M., Remillard R., Rappaport S.A., 1996, in `Supersoft X-ray Sources',
            J. Greiner (ed.), Lect. Notes in Phys. 472, 289 

\bibitem{} Boller T., Brandt W.N., Fink H.H., 1996, A\&A 305, 53 (BBF96)

\bibitem{} Boroson T.A., Green R.F., 1992, ApJS 80, 109

\bibitem{} Brandt W.N., Boller T., 1999, in `Structure and kinematics
              of quasar broad line regions', ASP conf. ser.,  
            C.M. Gaskell, W.N. Brandt, M. Dietrich et al. (eds), in press 

\bibitem{} Brandt W.N., Fabian A.C., Nandra K., Reynolds C.S., 
            Brinkmann W., 1994, MNRAS 271, 958   

\bibitem{} Brandt W.N., Fabian, A.C., Pounds K.A., 1996, MNRAS 278, 326   

\bibitem{} Brandt W.N., Mathur S., Elvis M., 1997, MNRAS 285, 25 

\bibitem{}Briel U., Aschenbach B., Hasinger G. et al., 1994, ROSAT
user's handbook, MPE: Garching

\bibitem{} Brinkmann W., Yuan W., Siebert J., 1997, A\&A 319, 413 

\bibitem{} Brinkmann W., Otani C., Wagner S., Siebert J., 1998, A\&A 330, 67

\bibitem{} Cappi M., Mihara T., Matsuoka M., et al., 1996, ApJ 458, 149 

\bibitem{} Ciliegi P., Maccacaro T., 1996, MNRAS 282, 477

\bibitem{} Comastri A., Fiore F., Guainazzi M., et al., 1998, A\&A 333, 31  

\bibitem{} Contini M., Viegas S.M., 1999, ApJ in press; astro-ph/9904226

\bibitem{} Dickey J.M., Lockman F.J., 1990, ARA\&A 28, 215

\bibitem{} Done C. Pounds K.A., Nandra K., Fabian A.C., 1995, MNRAS 275, 41 

\bibitem{} Fabian A.C., 1996, in MPE Report 263, H.U. Zimmermann, J. Tr\"umper,
                           H. Yorke (eds.), 403 

\bibitem{} Ferland G.J., 1993, University of Kentucky, Physics Department, 
Internal Report 

\bibitem{} Ferland G.J., Korista K.T., Verner D.A., et al., 1998, PASP 110, 761  

\bibitem{} Fink H.H., Walter R., Schartel N., Engels D., 1997, A\&A 317, 25 

\bibitem{} Fiore F., Elvis M., Mathur S., Wilkes B.J., McDowell J.C., 1993, ApJ 415, 129 

\bibitem{} Gaskell M., 1985, ApJ 291, 112

\bibitem{} George I.M., Mushotzky R., Turner T.J., et al., 1998, ApJ 509, 146

\bibitem{} Goodrich R.W., 1989, ApJ 342, 224

\bibitem{} Greiner J., 1996, in `Supersoft X-ray Sources', J. Greiner (ed.),
             Lect. Notes in Phys. 472, 285 

\bibitem{} Greiner J., Danner R., Bade N., et al.,  
                       1996, A\&A 310, 384 

\bibitem{} Grevesse N., Anders E., 1989, in `Cosmic Abundances of Matter',
                        C.J. Waddington (ed.), AIP 183, 1, 
                        New York: American Institute of Physics 

\bibitem{} Grupe D., 1996, PhD Thesis, Universit\"at G\"ottingen 

\bibitem{} Grupe D., Beuermann K., Thomas H.-C., Mannheim K., Fink H.H., 1998, A\&A 330, 25 

\bibitem{} Grupe D., Beuermann K., Mannheim K., Thomas H.-C., 1999a, A\&A, in press   

\bibitem{} Grupe D., Leighly K., Thomas, H.-C., 1999b, A\&A, submitted  

\bibitem{} Guainazzi M., Mihara T., Otani C., Matsuoka M., 1996, PASJ 48, 781

\bibitem{} Guainazzi M., Nicastro F., Fiore, et al., 1998a, MNRAS 301, L1

\bibitem{} Guainazzi M., Comastri A., Stirpe G.M., 1998b, A\&A 339, 327  

\bibitem{} Guilbert P.W., Fabian A.C., McCray R., 1983, ApJ 266, 466

\bibitem{} Halpern J.P., 1984, ApJ 281, 90

\bibitem{} Halpern J.P., Oke J.B., 1987, ApJ 312, 91

\bibitem{} Hayashida K., 1997, in `Emission Lines in Active Galaxies - New Methods
           and Techniques',
           B.M. Peterson, F.-Z. Cheng, A.S. Wilson (eds.), ASP conf. ser. 113, 40

\bibitem{} Iwasawa K., Brandt W.N., Fabian A.C., 1998, MNRAS 293, 251

\bibitem{} Kingdon J.B., Ferland G.J., 1999, ApJ 516, in press

\bibitem{}Komossa S., 1997, PhD Thesis, Ludw.-Max.-Univ. M\"unchen 

\bibitem{} Komossa S., Bade N., 1998, A\&A 331, L49

\bibitem{}Komossa S., Fink H., 1997a, A\&A 322, 719 

\bibitem{}Komossa S., Fink H., 1997b, A\&A 327, 483 

\bibitem{}Komossa S., Fink H., 1997c, A\&A 327, 555 

\bibitem{}  Komossa S., Fink H., 1997d, in `Accretion Disks -- New Aspects',
             E. Meyer-Hofmeister, H. Spruit (eds),  Lecture Notes in
             Physics 487, 250   

\bibitem{}  Komossa S., Fink H., 1997e, in `Emission Lines 
            in Active Galaxies: New Methods and Techniques', 
             B.M. Peterson, F.-Z. Cheng, A.S. Wilson (eds), ASP conf. ser. 113, 246 

\bibitem{}Komossa S., Greiner J., 1995, Astron Ges. Abstr. Ser. 11, 217

\bibitem{} Komossa S., Greiner J., 1999, in `High Energy Processes in Accreting Black Holes',
      J. Poutanen, R. Svensson (eds), ASP conf. ser. 161, 228  

\bibitem{} Komossa S., Janek M., 1999, in `Heating and acceleration
            in the universe',  H.Inoue, T.Ohashi,
            T.Takahashi (eds), Astron. Nachr. 320, in press; astro-ph/9907373

\bibitem{}Komossa S., Schulz H., 1997, A\&A 323, 31    


\bibitem{} Komossa S., Breitschwerdt D.,  
     Greiner J., Meerschweinchen J., 1999a, in `Astrophysical Dynamics', D. Berry,
     D. Breitschwerdt, A. da Costa, J. Dyson (eds),
     Ap\&SS, in press; astro-ph/9906330  

\bibitem{} Komossa S., Breitschwerdt D., Meerschweinchen J., 1999b, 
    in `Astrophysical Dynamics', D. Berry, D. Breitschwerdt, A. da Costa, J. Dyson (eds), 
    Ap\&SS, in press; astro-ph/9906377 

\bibitem{} Komossa S., Grupe D., Burwitz V., 2000, in `X-ray Astronomy 1999: 
             Stellar Endpoints, AGN and the X-ray Background',
             Astrophys. Lett. and Comm., in press  


\bibitem{} Krolik J.H., McKee C.F., Tarter C.B., 1981, ApJ 249, 422  

\bibitem{} Laor A., Fiore F., Elvis M., Wilkes B., McDowell J.C., 1994,
     ApJ 435, 611

\bibitem{} Laor A., Fiore F., Elvis M., Wilkes B., McDowell J.C., 1997,
     ApJ 477, 93

\bibitem{} Lawrence A., 1997, in `Emission Lines in Active Galaxies: New Methods and Techniques',
           B.M. Peterson, F.-Z. Cheng, A.S. Wilson (eds), ASP conf. ser. 113, 230

\bibitem{} Lawrence A., Watson M.G., Pounds K.A., Elvis M., 1985, MNRAS 217, 685

\bibitem{} Lawrence A., Elvis M., Wilkes B., McHardy I., Brandt N., 1997, MNRAS 285, 879 

\bibitem{} Leighly K.M., Mushotzky R.F., Yaqoob T., Kunieda K., Edelson R., 1996,
            ApJ 469, 14

\bibitem{} Leighly K.M., Kay L.E., Wills B.J., Wills D., Grupe D., 1997,
           ApJ 489, L137  

\bibitem{} Madau P., 1988, ApJ 327, 116

\bibitem{} Malkan M., 1986, ApJ 310, 679

\bibitem{} Mannheim K., Grupe D., Beuermann K., Thomas H.C., Fink H.H.,
               1996, in MPE Report 263, 
               H.U. Zimmermann, J. Tr\"umper, H. Yorke (eds), 471

\bibitem{} Marshall F.E., Holt S.S., Mushotzky R.F., Becker R.H., 1983, ApJ 269, L31

\bibitem{} Mathis J.S., Rumpl W., Nordsieck K.H., 1977, ApJ 217, 425 

\bibitem{} Mathur S., 1994, ApJ 431, L75 

\bibitem{} Matsuoka M., Piro L., Yamauchi M., Murakami T., 1990, ApJ 361, 440 

\bibitem{} McHardy I.M., Green A.R., Done C., et al., 1995, MNRAS 273, 549 

\bibitem{} Mihara T., Matsuoka M., Mushotzky R., et al., 1994, PASJ 46, L137 

\bibitem{} Miller P., Rawlings S., Saunders R., Eales S., 1992, MNRAS 254, 93

\bibitem{} Moran E.C., Halpern J.P., Helfand D.J., 1996, ApJS 106, 341 

\bibitem{} Nandra K., Pounds K.A., 1992, Nat 359, 215 

\bibitem{} Netzer H., 1990, in `Active Galactic Nuclei',
           Saas-Fee Lecture Notes 20, T.J.-L. Courvoisier, 
           M. Mayor (eds), Springer Verlag 

\bibitem{} Nicastro F., Fiore F., Perola G., Elvis M., 1999, ApJ 512, 184

\bibitem{} Osterbrock D.E., Dahari O., 1983, ApJ 273, 478

\bibitem{} Osterbrock D.E., Pogge R.W., 1985, ApJ 297, 166

 \bibitem{} Otani C., Kii T., Miya K., 1996,
        in MPE Report 263, H.U. Zimmermann, J. Tr\"umper, H. Yorke (eds), 491

\bibitem{} Pan H.C., Stewart G.C., Pounds K.A., 1990, MNRAS 242, 177 

\bibitem{} Pfeffermann E., Briel U.G., Hippmann H., et al., 
            1987, SPIE 733, 519

\bibitem{} Pounds K.A., Nandra K., Fink H.H., Makino F., 1994, MNRAS 267, 193

\bibitem{} Pounds K.A., Done C., Osborne J.P., 1995, MNRAS 277, L5   

\bibitem{} Puchnarewicz E.M., Mason K.O., Cordova F.A., et al., 1992, MNRAS 256, 589 

\bibitem{} Rafanelli P., Bonoli C., 1984, A\&A 131, 186 

\bibitem{} Reynolds C.S., Fabian, A.C., 1995, MNRAS 273, 1167  

\bibitem{} Reynolds C.S., Ward M.J., Fabian A.C., Celotti A., 1997, MNRAS 291, 493 

\bibitem{} Riley J.M., Warner P.J., Rawlings S., et al., 1988, MNRAS 236, 13p 

\bibitem{} Rodriguez-Pascual P.M., Mas-Hesse J.M., Santos-Lleo M., 1997, A\&A 327, 72 

\bibitem{} Rosenblatt E.I., Malkan M.A., Sargent W.L.W., Readhead A.C.S., 1992, ApJS 81, 59

\bibitem{} Schartel N., Walter R., Fink H., Tr\"umper J., 1996a, A\&A 307, 33

\bibitem{} Schartel N., Green P.J., Anderson S.F., et al., 1996b, MNRAS 283, 1015

\bibitem{} Schartel N., Komossa S., Brinkmann W., et al., 1997a, A\&A 320, 421

\bibitem{} Schartel N., Schmidt M., Fink H., Hasinger G., Tr\"umper J., 1997b, A\&A 320, 696

\bibitem{} Schwartz D.A., Zhao P., Remillard R., 1993, BAAS 25, 811

\bibitem{} Shakura N.I., Sunyaev R.A., 1973, A\&A 24, 337

\bibitem{} Shields J.C., Ferland G.J., Peterson B.M., 1995, ApJ 441, 507 

\bibitem{} Stephens S.A., 1989, AJ 97, 10 

\bibitem{}  Stocke J.T., Morris S.L., Gioia I.M., et al., 1991, ApJS 76, 813

\bibitem{}Tanaka Y., 1997, in `Accretion Disks -- New Aspects',
 E. Meyer-Hofmeister, H. Spruit (eds),
 Lecture Notes in Physics 487, 1

\bibitem{} Tr\"umper J., 1983, Adv. Space Res. 2, 241

\bibitem{} Turner T.J., Nandra K., George I.M., Fabian A.C., Pounds K.A., 1993, ApJ 419, 127

\bibitem{} Ulrich-Demoulin M.-H., Molendi S., 1996, ApJ 457, 77

\bibitem{} Vaughan S., Reeves J., Warwick R., Edelson R., 1999, MNRAS in press; Leicester Univ.
  preprint XRA 99/08 

\bibitem{} Walter R., Orr A., Courvoisier T.J.-L., Fink H.H., et al., 1994, A\&A 285, 119

\bibitem{} Wandel A., 1997, ApJ 490, L131   

\bibitem{} Wang T., Brinkmann W., Bergeron J., 1996, A\&A 309, 81 (WBB96)

\bibitem{} Wang T., Brinkmann W., Wamsteker W., Yuan W., 
             Wang Y.X., 1999, MNRAS, in press 

\bibitem{} Wisotzki L., Bade N., 1997, A\&A 320, 395 

\bibitem{} Xu D.W., Wei J.Y., Hu J.Y., 1999, ApJ 517, 622 

\bibitem{} Yuan W., 1998, PhD Thesis, TU M\"unchen 

\bibitem{} Zimmermann H.U., Becker W., Belloni T., et al., 1994,
               MPE Report 257

\end{thebibliography}
\end{document}